\documentclass[a4paper,11pt]{article}

\usepackage{jheppub} 

\usepackage[T1]{fontenc} 
\usepackage{slashed}
\usepackage{diagbox}
\usepackage{makecell}

\title{\boldmath Search for doubly charged scalars in type-II seesaw mechanism through photon fusion at the LHC}

\author[a]{Hang Zhou,}
\author[b,c]{Ning Liu}

\affiliation[a]{School of Microelectronics and Control Engineering, Changzhou University\\Changzhou, 213164, China}
\affiliation[b]{Physics Department and Institute of Theoretical Physics, Nanjing Normal University\\Nanjing, 210023, China}
\affiliation[c]{Nanjing Key Laboratory of Particle Physics and Astrophysics\\ Nanjing, 210023, China}

\emailAdd{zhouhang@cczu.edu.cn}
\emailAdd{liuning@njnu.edu.cn}

\abstract{Small neutrino masses can be generated through the well-known seesaw mechanisms, among which the type-II scenario predicts a triplet scalar with doubly charged components. Except for the Drell-Yan production at the Large Hadron Collider (LHC), the doubly charged scalars $\Delta^{\pm\pm}$ can also be produced through photon fusion along with ultraperipheral collision of protons, the outgoing protons from which can be detected by forward detectors at the LHC, providing a promising way to explore related new physics. We study the pair production through such processes at the 14 TeV LHC, focusing on the final states of $\mu^{+}\mu^{+}\mu^{-}\mu^{-}$ and $e^{+}e^{+}e^{-}e^{-}$ under normal (NH) and inverted hierarchy (IH) of the neutrino mass spectra, respectively. Promising sensitivity can be reached via our proposed search strategy. With luminosity of 36.1 fb$^{-1}$(100 fb$^{-1}$), $m_{\Delta}\sim430(520)$ GeV can be excluded at 95\% C.L. under the NH via $\mu^{+}\mu^{+}\mu^{-}\mu^{-}$ states searching, while the mass bound can be extended to 730(880) GeV under the IH via $e^{+}e^{+}e^{-}e^{-}$ states. The exclusion limits on $m_{\Delta}$ can be improved up to 1 TeV and even higher with integrated luminosity accumulated to 3 ab$^{-1}$.}

\begin{document}
\maketitle
\flushbottom

\section{Introduction}
\label{sec:intro}
Neutrino oscillation experiments strongly motivate small yet non-zero neutrino masses, providing clear evidence for new physics beyond the Standard Model (SM) of particle physics. One of the most well-known and widely-studied schemes to this problem is the Weinberg dimension-5 operator $\mathcal{L}\propto\ell_{L}HH\ell_{L}/\Lambda$~\cite{Weinberg:1979sa}, where $\ell_{L}$ and $H$ refer to the $SU(2)_{L}$ doublet lepton and the SM Higgs doublet, respectively. With a relatively high-scale cutoff $\Lambda$, tiny neutrino masses of Majorana nature can then be generated naturally after the electroweak symmetry spontaneous breaking (EWSB). Ultraviolet completion of the Weinberg operator at tree level exists in only three ways, generally known as the three types of seesaw mechanisms~\cite{minkowski1977,yanagida1979,ms1980,sv1980,mw1980,cl1980,lsw1981,ms1981,flhj1989,ma1998}, introducing to the SM right-handed neutrinos (type-I), $SU(2)_{L}$ triplet scalar (type-II) and $SU(2)_{L}$ triplet fermion (type-III), respectively. Different from type-I/III seesaw which mixes Dirac- and Majorana-type masses, the type-II scenario generates neutrino masses via Yukawa couplings of the triplet scalar $\Delta$ with the SM lepton doublets $m_{\nu}\sim Y_{\nu}v_{\Delta}$, where $Y_{\nu}$ is the neutrino Yukawa coupling and $v_{\Delta}$ the vacuum expectation value (VEV) of the neutral component of $\Delta$. A seesaw style thus appears after the EWSB considering the mixing between the SM Higgs and the triplet $v_{\Delta}\sim\mu v^{2}_{0}/m^{2}_{\Delta}$, with the triplet mass $m_{\Delta}$ being orders larger than the electroweak scale $v_{0}\sim246$ GeV. The dimensional mixing parameter $\mu$ in type-II seesaw is possible to be naturally small enough according to 't Hooft naturalness argument~\cite{Senjanovic:1978ev,tHooft:1979rat} so that in such a scenario, the Yukawa coupling $Y_{\mu}$ not only relates neutrino oscillation experiments data to collider searches through dileptonic decay of the new scalar, its mass $m_{\Delta}$ can also be low enough to be accessible at current collider experiments including the Large Hadron Collider (LHC).

Searches for the triplet scalar at colliders and studies on the relevant features have been extensively conducted~\cite{Ghosh:2017pxl,Crivellin:2018ahj,BhupalDev:2018tox,Altakach:2022hgn,Das:2023tna,Das:2024kyk,Arroyo-Urena:2025boh}, as its charged components, especially the doubly charged ones are predicted not only in the type-II seesaw, but also in a variety of other beyond SM models such as the left-right symmetric models~\cite{Pati:1974yy,Mohapatra:1974hk,Senjanovic:1975rk}, Zee-Babu model~\cite{Zee:1985id,Babu:1988ki}, Georgi-Machacek model~\cite{Chanowitz:1985ug,Georgi:1985nv}, etc. Pair production and associated production of these exotic scalars from Drell-Yan processes are commonly considered, and multileptonic final states are searched for as the most promising signals. The obtained experimental bounds for the doubly charged Higgs bosons are generally sensitive to different parametric regions of the model, mainly concerned with the triplet scalar VEV $v_{\Delta}$ and the mass spectrum of the charged and neutral components. It is the value of $v_{\Delta}$ that largely determines decay modes of the doubly charged scalars. Assuming $v_{\Delta}$ far less than $10^{-4}$ GeV which leads to dominant dileptonic decay channels $\Delta^{\pm\pm}\to\ell^{\pm}\ell^{'\pm}$, ATLAS collaboration derived a stringent lower limit for the doubly charged scalar mass at 95\% C.L. $m_{\Delta}>1080$ GeV~\cite{ATLAS:2022pbd}. With a larger $v_{\Delta}>10^{-4}$ GeV, the dileptonic decay channels are highly suppressed due to the increasing $v_{\Delta}$ and only diboson channels $\Delta^{\pm\pm}\to W^{\pm}W^{\pm}$ are relevant, relatively weaker bounds are obtained at $200\sim220$ GeV for a degenerate spectrum between the singly and doubly charged Higgs~\cite{ATLAS:2018ceg}. For a non-degenerate spectrum with a mass difference of up to 100 GeV, lower limits of 350 and 230 GeV are obtained for pair and associated production, respectively~\cite{ATLAS:2021jol}. Although most current experimental searches assumed degeneracy or a tiny mass splitting, it is allowed to be as large as $\Delta m\sim40$ GeV for hundreds-GeV triplet Higgs if considering the electroweak precision data (EWPD) due to the fact that heavy scalars contribution to the oblique parameter $T$ is compensated by the mass difference among the triplet components~\cite{Melfo:2011nx,Chun:2012jw}. In addition, for a moderate mass separation $\Delta m\gtrsim5$ GeV or even larger splitting, cascade decay channels $\Delta^{\pm\pm}\to\Delta^{\pm}W^{\pm*}\to\Delta^{0}W^{\pm*}W^{\pm*}$ become dominant over the dileptonic and dibosonic ones if $v_{\Delta}$ is not very far from $10^{-4}$ GeV by orders (either larger or smaller)~\cite{Melfo:2011nx}. Hence, when the decay of the new scalars is dominated by different channels, quite distinct signatures appear in collider phenomenology. And due to the large hadronic background at the LHC, a tiny $v_{\Delta}$ is typically assumed for a dominant dileptonic mode in experimental searches as well as phenomenological studies.

While the above-mentioned conclusions on experimental bounds are mostly drawn by taking into consideration of Drell-Yan production processes of the triplet scalars, initial photon fusion from elastic collisions of protons, although receiving less attention, becomes increasingly attractive recently as the forward detectors have been launched at the LHC, including the CMS-TOTEM Precision Proton Spectrometer (CT-PPS)~\cite{CTPPS2014} and the ATLAS Forward Proton detector (AFP)~\cite{AFP2015}. These forward physics facilities (FPF) are located close to the colliding beams and around 220 meters from the collision point. With the FPFs installed successfully, new windows are opened to the elastic events at the LHC since the forward detectors are designed mainly to identify the unsuccessfully colliding protons. Those protons go through what are generally known as the ultraperipheral collisions (UPC) and remain as intact ones when reaching the forward detectors. Along with the UPC processes, as the electromagnetic field around the fast-moving protons can be approximated to on-shell photons (equivalent photon approximation, EPA), initial photon fusion can occur accompanying the elastic collisions between protons, leading to pair production of charged particles such as the doubly charged scalars. Although it has been shown that contribution from photon-fusion is less than that of neutral current Drell-Yan production~\cite{Han:2007bk,Fuks:2019clu}, the elastic kind of collisions will leave unique signatures of two unharmed protons reaching the forward detectors, forming a special topology with rapidity gaps between those forward protons and the central particles. These features make the photon fusion in UPC processes a new, promising way to search for new physics in the forward regions, which was not even possible before the launch of forward facilities. Utilizing forward proton-tagging, an increasing number of phenomenological studies have been conducted for probing BSM particles at the LHC, including the supersymmetric (SUSY) dark matter candidates~\cite{Harland-Lang:2018hmi,Beresford:2018pbt}, quasistable or nearly-degenerate charginos in specific SUSY scenarios~\cite{Godunov:2019jib,Zhou:2022jgj,Zhou:2024fjf} and multiple charged scalars in seesaw and left-right symmetric models~\cite{Babu:2016rcr,Duarte:2022xpm,Duarte:2024zeh}.

In this paper, we will propose a search strategy for probing degenerate doubly-charged scalars in the type-II seesaw model via tagging forward protons at the LHC, within the context of both normal and inverted hierarchies of neutrino mass spectra. In the next section, we concisely introduce the multiple scalars of the type-II seesaw and their production and decay at the hadron collider, as well as the connection to neutrino physics. In Section III, details on the signals and simulation will be presented. Section IV is about our proposed search strategies and results. Finally, we draw the conclusion in Section V.

\section{Doubly charged scalars in type-II seesaw model}
\label{sec:model}

The seesaw mechanism of type-II for neutrino mass generation can be realized in a simple way by extending the SM with a complex scalar triplet lying in the adjoint representation of the weak group $SU(2)_{L}$,
\begin{align}
\Delta=
\begin{pmatrix}
\frac{\Delta^{+}}{\sqrt{2}} && \Delta^{++} \\
\Delta^{0} && -\frac{\Delta^{+}}{\sqrt{2}}\,
\end{pmatrix},
\end{align}
with hypercharge $Y_{\Delta}=1$ as in the convention for the formula $Q=T_{3}+Y$. A gauge-invariant and most general renormalizable Lagrangian for the type-II seesaw scalar sector can be expressed as
\begin{align}
\mathcal{L}\subset(D_{\mu}H)^{\dagger}(D^{\mu}H)+\text{Tr}(D_{\mu}\Delta)^{\dagger}(D^{\mu}\Delta)-V(H,\,\Delta),
\end{align}
where $H$ is the SM Higgs doublet $H=\left(\phi^{+},\,\phi^{0}\right)^{\text{T}}$ and $V(H,\Delta)$ stands for the scalar potential including contributions from the SM Higgs and the triplet scalar,
\begin{align}
\label{pot}
V(H,\Delta)={}&-m^{2}_{H}H^{\dagger}H+m^{2}_{\Delta}\text{Tr}[\Delta^{\dagger}\Delta]+[\mu H^{T}i\sigma_{2}\Delta^{\dagger}H+\text{h.c.}]+\frac{\lambda}{4}(H^{\dagger}H)^{2}\notag\\
{}&+\lambda_{1}(H^{\dagger}H)\text{Tr}[\Delta^{\dagger}\Delta]+\lambda_{2}[\text{Tr}(\Delta^{\dagger}\Delta)]^{2}+\lambda_{3}\text{Tr}[(\Delta^{\dagger}\Delta)^{2}]+\lambda_{4}H^{\dagger}\Delta^{\dagger}\Delta H\,,
\end{align}
in which $m^{2}_{H,\Delta}$ are mass parameters and $\lambda$, $\lambda_{1\sim4}$ are quartic couplings that can be taken as real numbers without loss of generality. $\mu$ is the coupling for trilinear terms, which may violate lepton number conservation.

Besides the kinetic terms and scalar potential, type-II seesaw introduces an additional Yukawa interaction between the triplet scalar and the SM lepton doublets $L=(\nu_{\ell}\,,\ell)^{T}$ for neutrino mass generation,
\begin{align}
\mathcal{L}_{Y_{\Delta}}=-y_{\Delta}L^{T}Ci\sigma_{2}\Delta L+\text{h.c.}\,,
\end{align}
where $y_{\Delta}$ is the Yukawa coupling and $C$ the charge conjugation operator. Once the doublet and triplet scalars develop non-vanishing vacuum expectation values after the spontaneous electroweak symmetry breaking as $v_{0}$ and $v_{\Delta}$, respectively, the seesaw mechanism works to generate Majorana masses of neutrinos
\begin{align}
\mathcal{M}_{\nu}=\sqrt{2}Y_{\Delta}v_{\Delta}\,,
\end{align}
where the complex symmetric matrix $\mathcal{M}_{\nu}$ can be diagonalized using the PMNS mixing matrix $U$ as $\mathcal{M}_{\nu}=U^{*}m_{\nu}U^{\dagger}$. Diagonal $m_{\nu}$ incorporates three physical neutrino masses mixed by three flavors through the mixing matrix, entries of which can be largely determined by various neutrino oscillation experiments except for the minimal neutrino mass $\nu_{1/3}$, corresponding to normal hierarchy (NH)/inverted hierarchy (IH), and two Majorana phases. As the minimalization of the scalar potential Eq~\eqref{pot} leads to $v_{\Delta}\approx\mu v^{2}_{0}/m^{2}_{\Delta}$ with the electroweak vacuum $v^{2}=v^{2}_{0}+v^{2}_{\Delta}\approx246^{2}\,\text{GeV}^{2}$, one can infer that a large $m_{\Delta}$ or a small $\mu$ can both induce a small $v_{\Delta}$, giving tiny neutrino masses in a seesaw way. Besides, triplet scalar VEV contributes to radiative corrections for $\rho$ parameter in electroweak precision observables as a result of modifications to $W$ and $Z$ masses, leading to the prediction $\rho\approx1-2v^{2}_{\Delta}/v^{2}$. Given the current results for global fit $\rho=1.00031\pm0.00019$~\cite{ParticleDataGroup:2024cfk}, an upper limit for $v_{\Delta}$ can be obtained around 2.8 GeV at $3\sigma$ C.L., far less than the doublet VEV $v_{0}$. Since the non-diagonal Yukawa matrix $y_{\Delta}$ gives rise to lepton flavor violating processes including $\mu^{-}\to e^{+}e^{-}e^{-}$ and $\mu^{-}\to e^{-}\gamma$, upper bounds for the branching ratios of these rare decays~\cite{SINDRUM:1987nra,MEG:2016leq} can then be translated into a bound from below for $v_{\Delta}$ depending on $m_{\Delta^{\pm\pm}}$: $v_{\Delta}\gtrsim10^{-9}\text{GeV}\times(1\text{TeV}/m_{\Delta^{\pm\pm}})$~\cite{Dinh:2012bp,Ashanujjaman:2021txz}.

As stated briefly above in Sec.~\ref{sec:intro}, decay modes of the doubly charged scalars depend on the triplet VEV $v_{\Delta}$ and their mass spectrum. Decay width for the leptonic decay modes can be given as~\cite{Melfo:2011nx,Chun:2003ej,FileviezPerez:2008jbu,Mandal:2022zmy}:
\begin{align}
\label{lepbr}
\Gamma(\Delta^{\pm\pm}\to\ell^{\pm}_{i}\ell^{\pm}_{j})=\frac{m_{\Delta^{\pm\pm}}}{8\pi(1+\delta_{ij})}\left|\frac{\mathcal{M}^{ij}_{\nu}}{v_{\Delta}}\right|^{2},
\end{align}
which, if $v_{\Delta}<10^{-4}$ GeV, dominate over the gauge bosonic ones:
\begin{align}
\Gamma(\Delta^{\pm\pm}\to W^{\pm}W^{\pm})=\frac{g^{4}v^{2}_{\Delta}}{8\pi m_{\Delta^{\pm\pm}}}\sqrt{1-\left(\frac{2m_{W}}{m_{\Delta^{\pm\pm}}}\right)^{2}}\left[\left(\frac{m^{2}_{\Delta^{\pm\pm}}}{2m^{2}_{W}}-1\right)^{2}+2\right]\,,
\end{align}
where $i$, $j$ stand for lepton flavors, $\delta_{ij}$ the Kronecker delta symbol and $g$ the weak gauge coupling. Cascade decays are possible kinematically only if $\Delta m>0$ and are dominant for a sufficiently large value of $\Delta m$, which is beyond the scope of the present work focusing on the case of a degenerate mass spectrum. Meanwhile, considering the constraints from electroweak precision measurements and lepton flavor violating decays discussed above, we can safely assume a triplet VEV much smaller than $10^{-4}$ GeV so that only leptonic decay modes will be relevant in our following analysis of the collider search for the charged Higgs bosons.

\section{Signal and background}
\label{sec:signal}

\begin{figure}[t]
\centering
\includegraphics{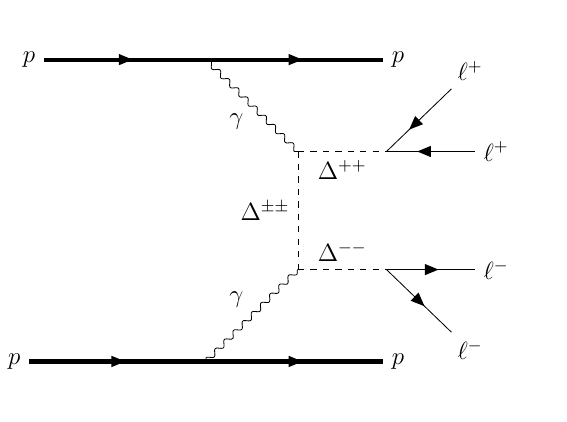}
\caption{Feynman diagram for the signal process of fully leptonic channel from doubly charged scalars pair production through elastic EPA photon fusion at the LHC.}
\label{fig:sigFD}
\end{figure}

\begin{figure}[t]
\centering
\begin{minipage}{0.38\linewidth}
  \centerline{\includegraphics[scale=0.78]{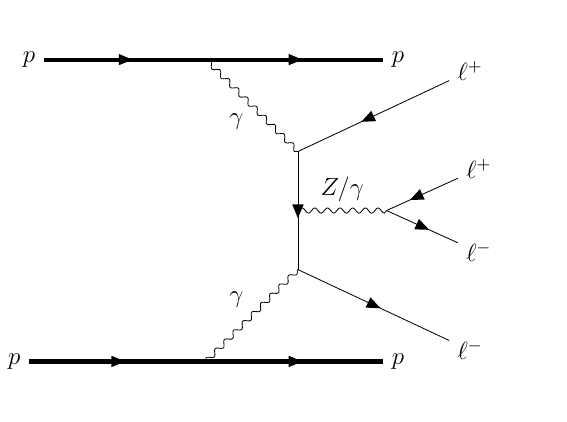}}
  \centerline{(a)}
\end{minipage}
\qquad\qquad
\begin{minipage}{0.38\linewidth}
  \centerline{\includegraphics[scale=0.78]{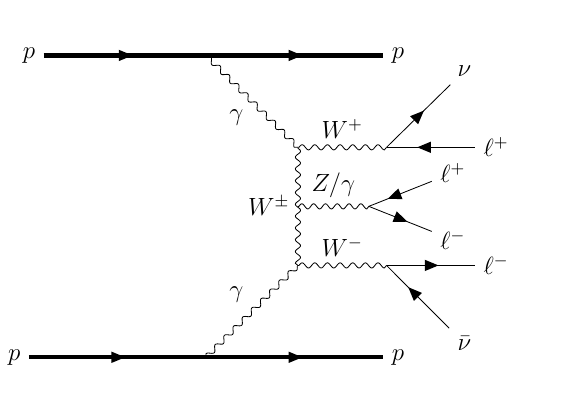}}
  \centerline{(b)}
\end{minipage}
\\[12pt]
\begin{minipage}{0.38\linewidth}
  \centerline{\includegraphics[scale=0.78]{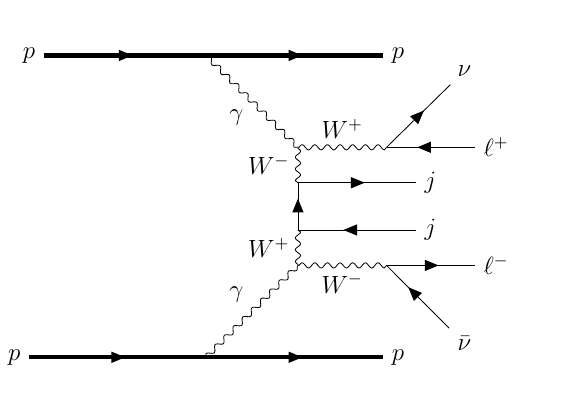}}
  \centerline{(c)}
\end{minipage}
\qquad\qquad
\begin{minipage}{0.38\linewidth}
  \centerline{\includegraphics[scale=0.78]{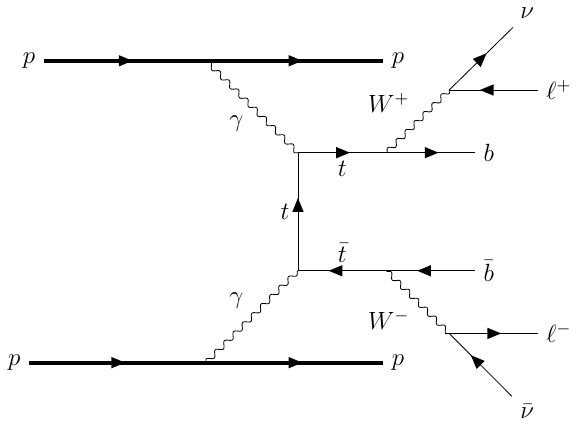}}
  \centerline{(d)}
\end{minipage}
\caption{Feynman diagrams of the Standard Model background for fully leptonic channels through elastic EPA photon fusion at the LHC. $t$-channel $Z/\gamma$-mediating diagrams in (a) and (b) are not shown but taken into consideration in our analysis.}
\label{fig:bkgFD}
\end{figure}

Different from the well-studied Drell-Yan production for the doubly charged scalars introduced in Sec.~\ref{sec:intro}, we study in the present paper a search strategy through photon fusion pair production $\gamma\gamma\to\Delta^{\pm\pm}\Delta^{\mp\mp}$, where the initial photons come from equivalent photon approximation (EPA) alongside the ultraperipheral collision of protons at the LHC. Two colliding protons remain intact in the final states to be tagged by the forward detectors. This way of production arises from the fully elastic channel whose total cross section can be written as
\begin{align}
\sigma_{pp\to p(\gamma\gamma\to\Delta^{++}\Delta^{--})p}=\int\int\sigma_{\gamma\gamma\to\Delta^{++}\Delta^{--}}\gamma^{el}_{1}(z_{1})\gamma^{el}_{2}(z_{2})dz_{1}dz_{2}
\end{align}
where $\gamma^{el}_{1/2}(z_{1/2})$ are $\gamma$-PDFs, distribution functions of the elastic equivalent photon in a proton with $z_{1/2}$ being momentum fractions of EPA photons. Elastic $\gamma$-PDFs indicate the probabilities a proton radiates elastically a photon while itself undamaged, analytic expressions of which can be found in~\cite{Budnev:1975poe,Kniehl:1990iv}. The convolution integral over the momenta should not be taken over the whole range of parametric space, as the forward detectors measure final protons with various efficiencies depending on the proton energy loss $\xi\equiv1-E_{\text{out}}/E_{\text{in}}$, with $E_{\text{in}/\text{out}}$ referring to energies of the ingoing/outgoing protons. These effects are taken into account by translating the tagging efficiencies for protons to the ones for the EPA photons, which we will discuss below.

\begin{figure}[t]
\centering
\includegraphics[scale=0.8]{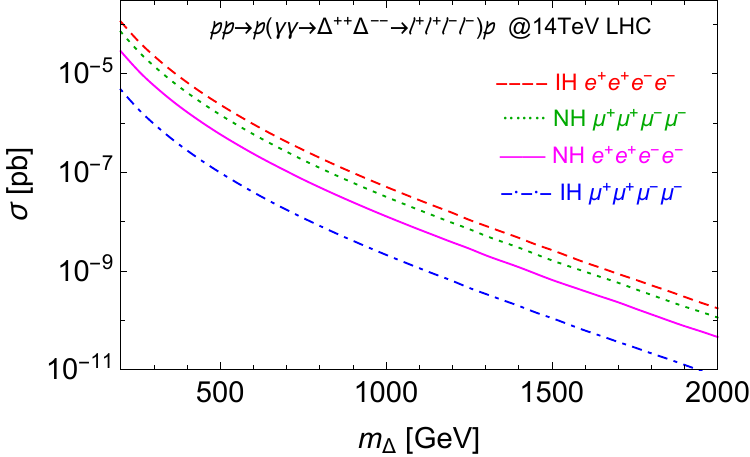}
\caption{Total cross sections of $pp\to p(\gamma\gamma\to\ell^{+}\ell^{+}\ell^{-}\ell^{-})p$ with 4-electron and 4-muon final states for both normal and inverted hierarchies of neutrino mass spectra.}
\label{fig:xs}
\end{figure}

To realize a better sensitivity at the hadronic environment at the LHC, we assume $v_{\Delta}<10^{-4}$ GeV as discussed in the previous section and decay the doubly charged Higgs bosons into pairs of leptons ($e^{\pm}$ or $\mu^{\pm}$). Such a process through EPA photon fusion can then be expressed as
\begin{align}
\label{proc:sig}
pp\to p(\gamma\gamma\to\Delta^{++}\Delta^{--}\to\ell^{+}_{1}\ell^{+}_{2}\ell^{-}_{3}\ell^{-}_{4})p\,,
\end{align}
where $\ell_{1,2,3,4}$ can be different combinations of lepton flavors, corresponding Feynman diagrams are displayed in Figure.~\ref{fig:sigFD}, drawn using the TikZ-Feynman package~\cite{Ellis:2016jkw}. According to the expression of decay width Eq.\eqref{lepbr}, although the nature of neutrino mass is still unknown which is sensitive to the neutrinoless double beta decay experiment, the oscillation data can to a large extent fix some elements of the neutrino mass matrix $\mathcal{M}_{\nu}$ and then the branching ratios of certain leptonic decay channels. In our following simulation, we adopt the best-fit values for neutrino mixing parameters for the NH and IH cases~\cite{Esteban:2024eli,nufit}
\begin{align}
\label{pmns:NH}
\text{NH}: \quad{}&\Delta m^{2}_{21}=7.49\times10^{-5}\text{eV}^{2},\,\,\Delta m^{2}_{31}=2.513\times10^{-3}\text{eV}^{2},\notag\\ {}&\sin^{2}\theta_{12}=0.308,\,\,\sin^{2}\theta_{23}=0.470,\,\,\sin^{2}\theta_{13}=0.02215,\,\, \delta_{CP}=212^{\circ},\\
\label{pmns:IH}
\text{IH}:\quad{}&\Delta m^{2}_{21}=7.49\times10^{-5}\text{eV}^{2},\,\,\Delta m^{2}_{31}=-2.484\times10^{-3}\text{eV}^{2},\notag\\ {}&\sin^{2}\theta_{12}=0.308,\,\,\sin^{2}\theta_{23}=0.550,\,\,\sin^{2}\theta_{13}=0.02231,\,\, \delta_{CP}=274^{\circ}.
\end{align}
Note that two Majorana phases are assumed vanishing and the lightest neutrino mass is adopted as $10^{-4}$ eV. With these best-fit values input, the leading decay channel is $\Delta^{\pm\pm}\to\mu^{\pm}\tau^{\pm}$\,(BR$\sim35\%$) for NH case and $\Delta^{\pm\pm}\to e^{\pm}e^{\pm}$\,(BR$\sim47\%$) for IH case~\cite{Li:2023ksw}. In consideration of better detection efficiencies at the LHC experiments for electrons and muons, we choose as our signal where both scalars decay into $\ell_{1\sim4}=\mu$ (BR$\sim25\%$ for $\Delta^{\pm\pm}\to\mu^{\pm}\mu^{\pm}$ under the above assumptions for neutrino mixing~\cite{Li:2023ksw}) for NH and $\ell_{1\sim4}=e$ for IH. Figure.~\ref{fig:xs} displays cross sections for these processes versus the triplet scalar mass ranging from 200 GeV to 2 TeV, in which the difference of branching ratios of $\Delta$ decaying into different lepton flavors also manifests itself as discussed above: for NH, the muon channel surpasses the electron one and vice versa for IH.

Standard Model background corresponding to the above signals includes direct production of charged leptons through EPA photon fusion associated with a $Z/\gamma$ decaying into a lepton pair (Figure.~\ref{fig:bkgFD}(a)), as well as a $W$ pair production through the photon fusion associated with a $Z/\gamma$ (Figure.~\ref{fig:bkgFD}(b)), with two intact protons in the final state,
\begin{align}
\label{proc:bkg12}
pp\to{}&p(\gamma\gamma\to\ell^{+}\ell^{-}Z/\gamma\to\ell^{+}\ell^{-}\ell^{+}\ell^{-})p\,,\\
pp\to{}&p(\gamma\gamma\to W^{+}W^{-}Z\to\ell^{+}\nu\ell^{-}\bar{\nu}\ell^{+}\ell^{-})p\,.
\end{align}
Note that in the above diagrams, the mediating gauge boson $Z/\gamma$ is shown going through $s$-channel, while a $t$-channel contribution is also considered but not shown in the figures for simplicity. Similar to the process in Figure.~\ref{fig:bkgFD}(b), associated production of two light jets along with the photon fusion into two $W$ bosons can also mimic the signal when the jets are misidentified as leptons, so the following process is also considered as one of our backgrounds (Figure.~\ref{fig:bkgFD}(c))
\begin{align}
\label{proc:bkg3}
pp\to&p(\gamma\gamma\to W^{+}W^{-}jj\to\ell^{+}\nu\ell^{-}\bar{\nu}jj)p\,.
\end{align}
Another background is top quark pair production through EPA photon fusion, followed by the top leptonic decay (Figure.~\ref{fig:bkgFD}(d)),
\begin{align}
\label{proc:bkg4}
pp\to&p(\gamma\gamma\to t\bar{t}\to b\ell^{+}\nu\bar{b}\ell^{-}\bar{\nu})p\,,
\end{align}
which also contaminates the signal with $b$-jets misidentification as leptons.

\section{Search strategies and results}
\label{sec:search}

\begin{table}
\centering
\begin{tabular}{|c|*{5}{c|}}
\hline
\,$E_{\gamma}$\,(GeV)\, & \,(0,100]\, & \,(100,120]\, & \,(120,150]\, & \,(150,400]\, & \,(400,$+\infty$)\, \\ \hline
Eff. & 0 & 50\% & 70\% & 90\% & 80\% \\ \hline
\hline
\end{tabular}
\caption{Acceptance rates for initial photons for different ranges of energies, which are equivalent to tagging efficiencies for the outgoing protons corresponding to their energy losses \cite{CTPPS2014,AFP2015}.}
\label{tab:protonrates}
\end{table}

To arrive at an effective search strategy for the fully leptonic channel from the doubly charged scalar at the LHC, we perform simulations using \textsc{MadGraph5\_aMC@NLO} (v3.5.6)~\cite{Alwall2014} for parton-level events generation and calculations for cross sections. Parton showering and detector simulation are realized by \textsc{Pythia-8.2}~\cite{Sjostrand:2014zea} and \textsc{Delphes-3.5.0}~\cite{deFavereau:2013fsa} embedded in \textsc{CheckMATE2}~\cite{Dercks2017} for further analysis and event selection. For model files, we use the \textsc{Typeiiseesaw} Universal \textsc{Feynrules} Output (UFO) libraries that are developed in~\cite{Fuks:2019clu}. MMHT+LUXqed NLO PDF sets~\cite{Buckley:2014ana,Manohar:2016nzj,Manohar:2017eqh,Harland-Lang:2019pla} are used for event generation of photon fusion. More discussion about $\gamma$-PDFs can be found below in the end of this section when we present our results and compare them with ones using other $\gamma$-PDF.

\begin{figure}
\centering
\begin{minipage}{0.38\linewidth}
  \centerline{\includegraphics[scale=0.38]{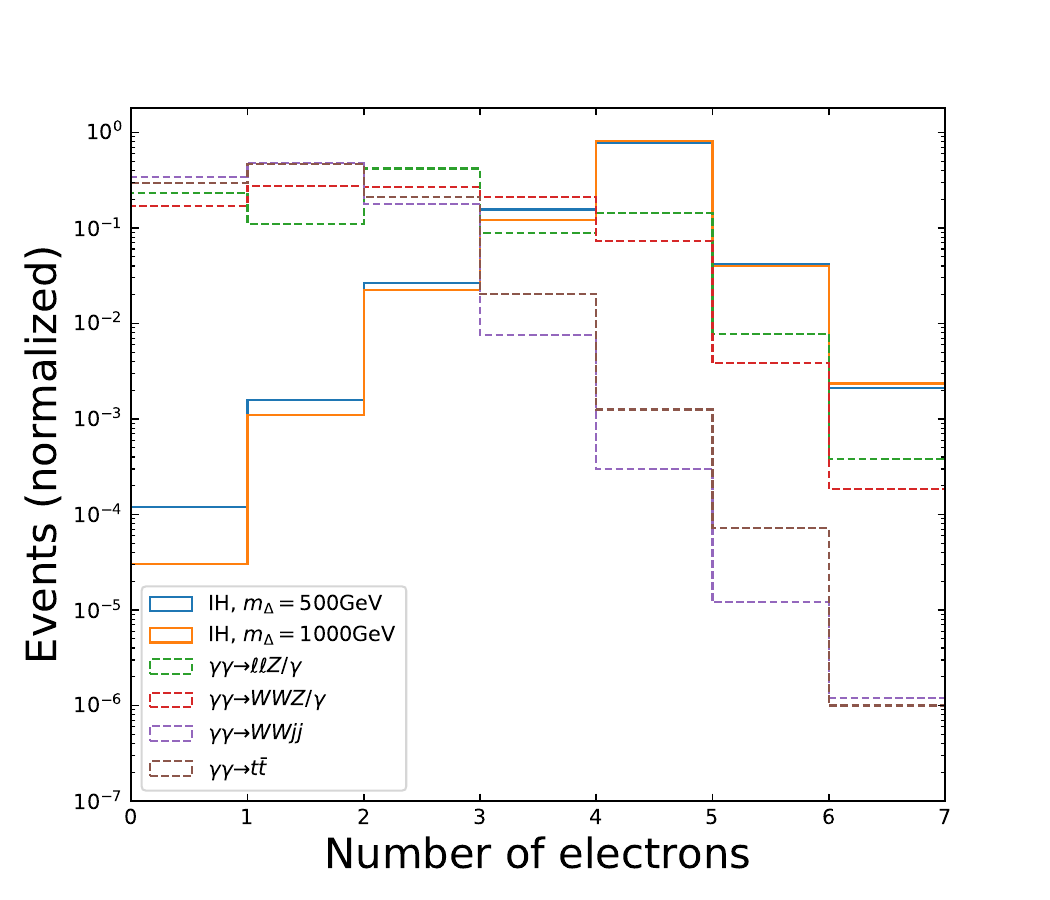}}
  \centerline{(a)}
\end{minipage}
\qquad\qquad
\begin{minipage}{0.38\linewidth}
  \centerline{\includegraphics[scale=0.38]{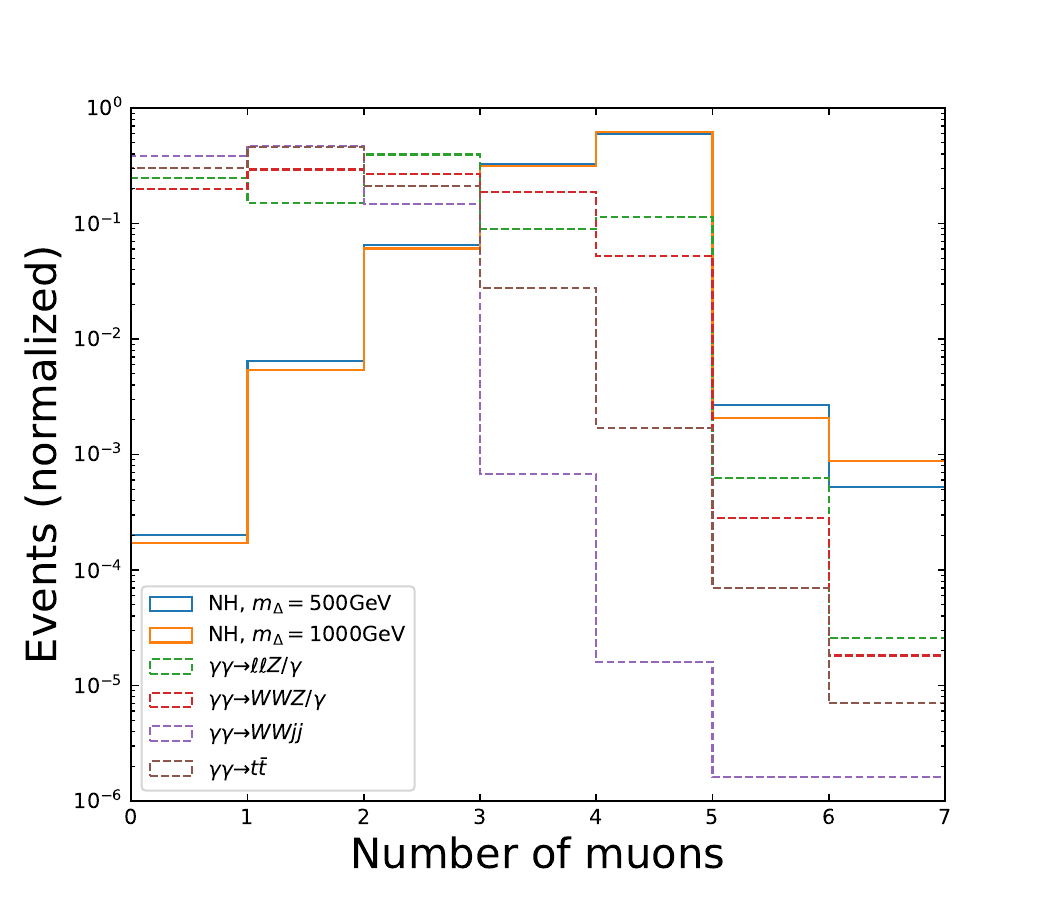}}
  \centerline{(b)}
\end{minipage}
\begin{minipage}{0.38\linewidth}
  \centerline{\includegraphics[scale=0.38]{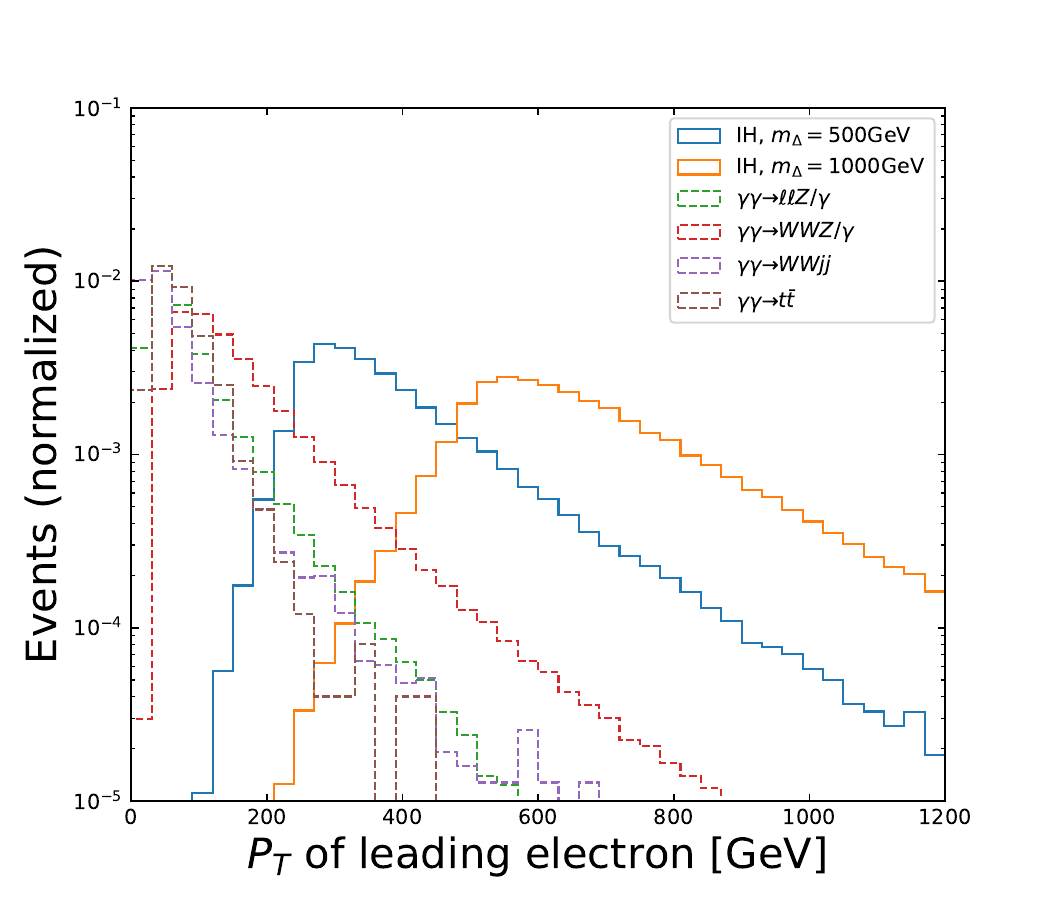}}
  \centerline{(c)}
\end{minipage}
\qquad\qquad
\begin{minipage}{0.38\linewidth}
  \centerline{\includegraphics[scale=0.38]{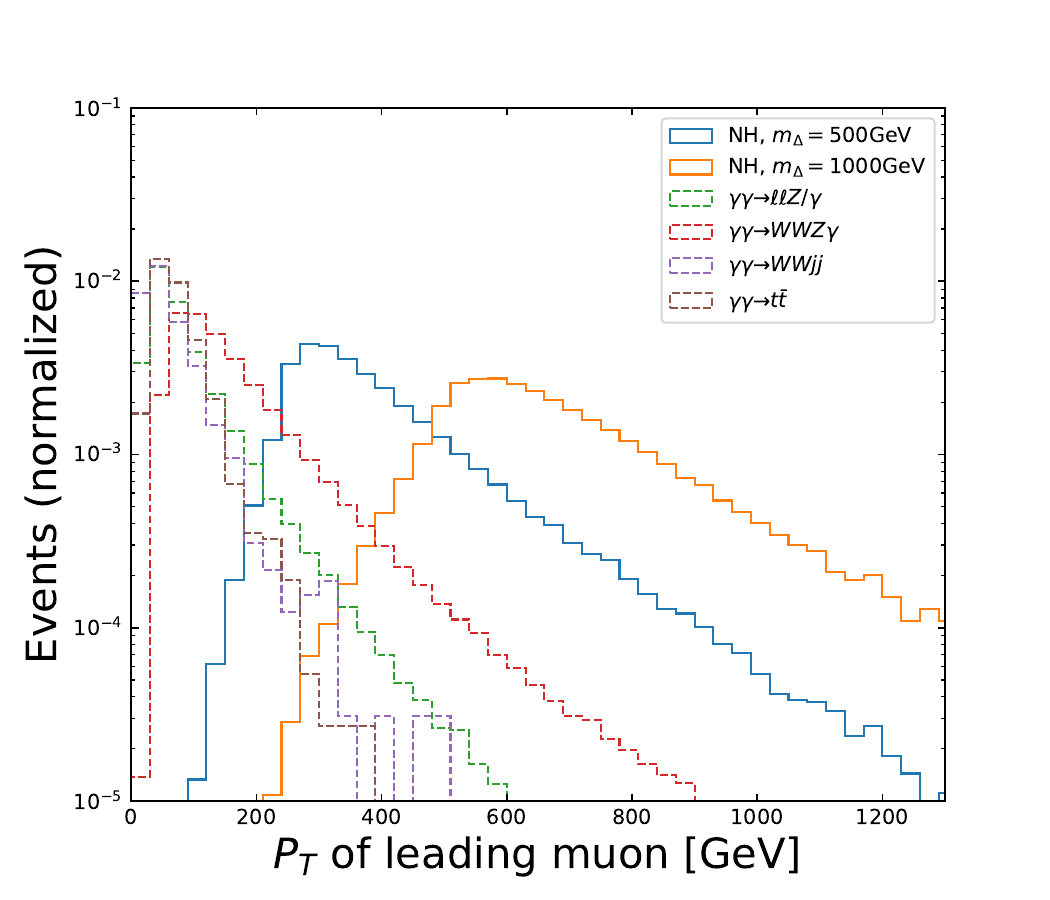}}
  \centerline{(d)}
\end{minipage}
\begin{minipage}{0.38\linewidth}
  \centerline{\includegraphics[scale=0.38]{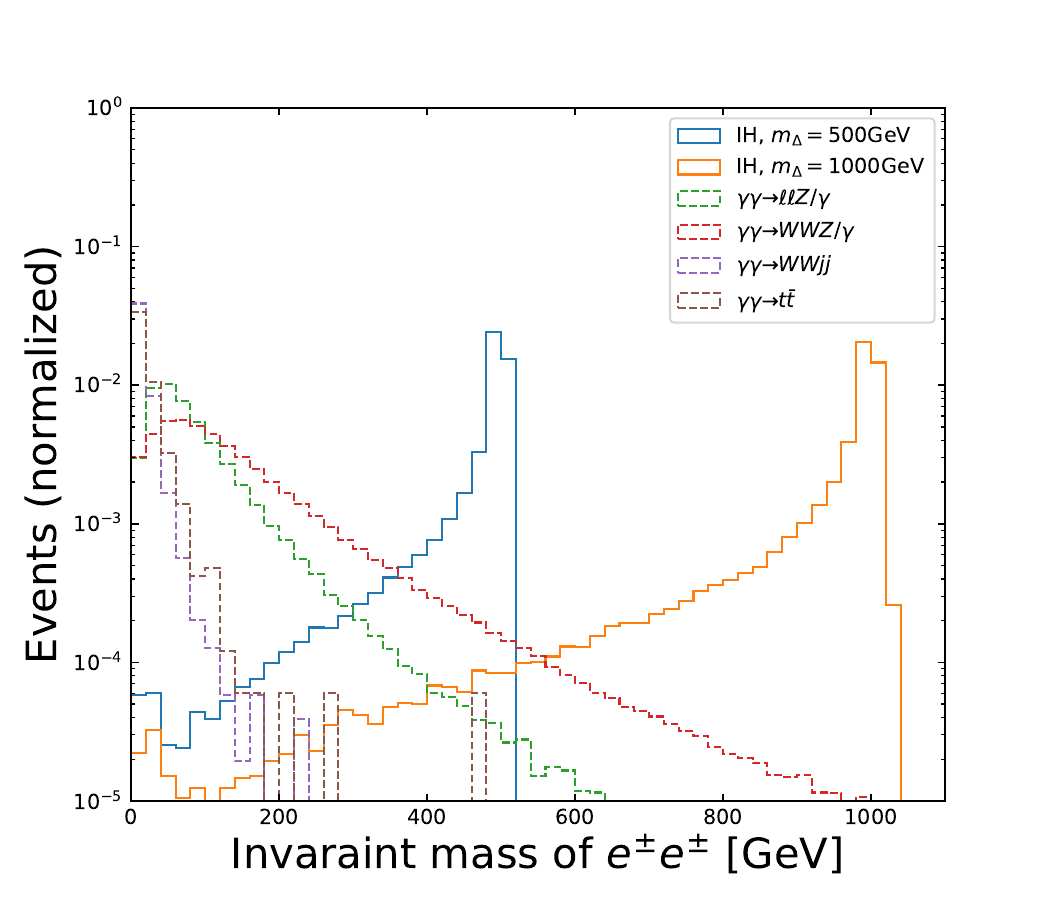}}
  \centerline{(e)}
\end{minipage}
\qquad\qquad
\begin{minipage}{0.38\linewidth}
  \centerline{\includegraphics[scale=0.38]{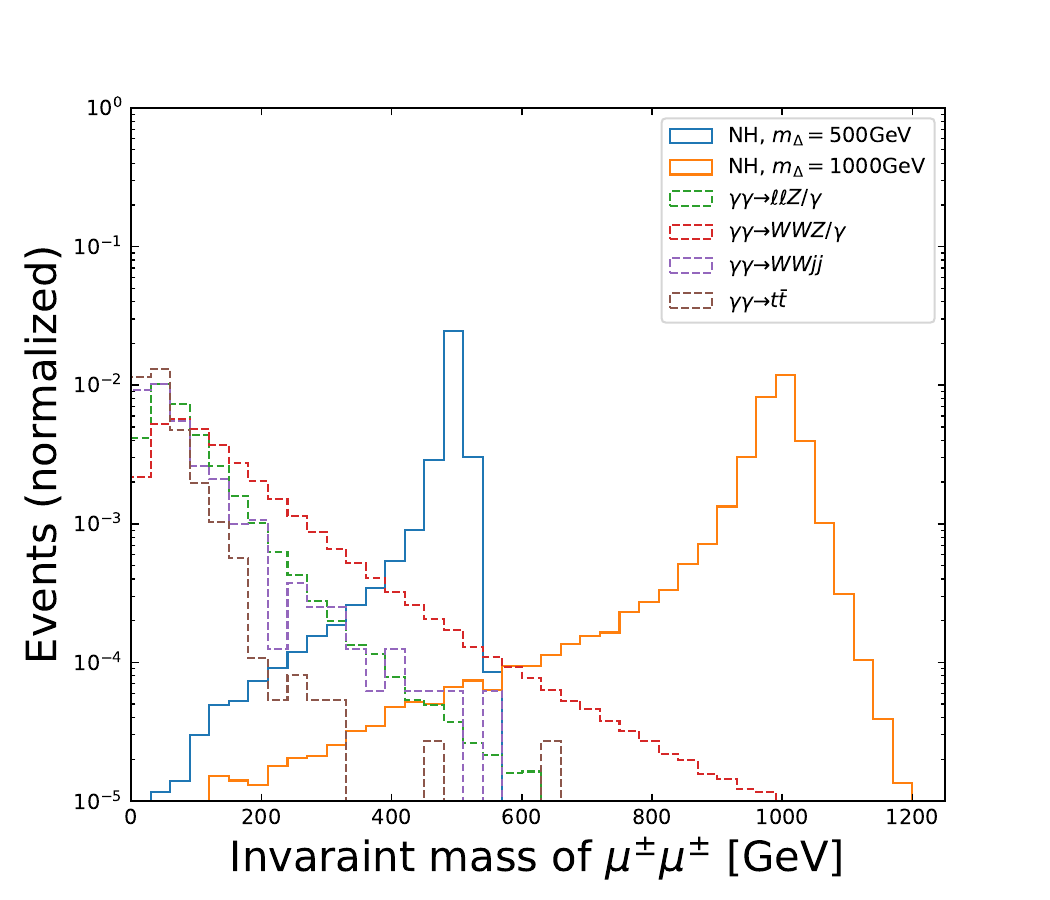}}
  \centerline{(f)}
\end{minipage}
\caption{Kinematic distributions of electron (muon) number, transverse momentum of the leading electron (muon) and invariant mass of same-sign electrons (muons) from EPA photon fusion at the LHC: $pp\to p(\gamma\gamma\to\Delta^{++}\Delta^{--}\to\ell^{+}\ell^{+}\ell^{-}\ell^{-})p$\,, for inverted (normal) hierarchy spectrum as signal, as well as for four SM background processes at the LHC. Histograms for the signals are shown in solid lines and those for backgrounds are shown in dashed lines.}
\label{fig:dist}
\end{figure}

One of the most characteristic features of the signal is the final two intact protons and, as discussed in Section~\ref{sec:signal}, they are not 100\% detectable by the forward detectors at the LHC, but rather measured at some rates depending on their energy losses. In view of the elastic nature of the ultraperipheral collision, the energies of EPA photons $E_{\gamma}$ emitted from the protons can be considered equal to the energy losses. For $E_{\gamma}$ from 100 GeV to 1 TeV, the detection rates for final protons approximate to 100\% at the 13 TeV LHC, which can be translated to the energy loss $\xi$ of (0.015, 0.15) defined in Section~\ref{sec:signal}~\cite{CTPPS2014,AFP2015}. In our simulation, we adopt more conservative detection rates (Table.~\ref{tab:protonrates}) since lower efficiencies around 90\% are generally indicated by phenomenological studies~\cite{Beresford:2018pbt}. These values of proton tagging rates are then applied to the generated events, leaving only part of the events for further selection. This procedure can be viewed as a pre-selection in which the four-momenta of final protons are smeared with a 5\%-width Gaussian function prior to the tagging rates application to realize the forward detector simulation. We use \textsc{PYLHE}~\cite{pylhe} package to process the MG5-generated events and to do the smearing.

\begin{table}
\centering
\begin{tabular}{|c|c|c|}
\hline
\diagbox{Cuts}{Signal} & $e^{+}e^{+}e^{-}e^{-}$, IH &  $\mu^{+}\mu^{+}\mu^{-}\mu^{-}$, NH  \\ \hline
Pre-selection & \multicolumn{2}{|c|}{Two outgoing protons tagged}  \\ \hline
Cut-1 & $e^{+}e^{+}e^{-}e^{-}$ & $\mu^{+}\mu^{+}\mu^{-}\mu^{-}$ \\ \hline
Cut-2 & \multicolumn{2}{|c|}{Leading $e/\mu$ $p_{T}>50$}  \\ \hline
Cut-3 & $m(e^{\pm}e^{\pm})\in[m_{\Delta}-25,m_{\Delta}+15]$ & $m(\mu^{\pm}\mu^{\pm})\in[m_{\Delta}-20,m_{\Delta}+20]$ \\
\hline
\end{tabular}
\caption{Event selections for signal processes with final states of $e^{+}e^{+}e^{-}e^{-}$ and $\mu^{+}\mu^{+}\mu^{-}\mu^{-}$, corresponding to IH and NH, respectively. Units of $p_{T}$ and masses are GeV which are omitted in the table for simplicity.}
\label{tab:selection}
\end{table}

Events surviving the pre-selection then go through further cuts based on different features of kinematic distributions for the signal and background. We present for illustration the histograms of three typical kinematic variables for signal and background events in Figure.~\ref{fig:dist}, including the number of electrons (muons), transverse momentum of the leading electron (muon) and the invariant mass of same-sign electrons (muons) for inverted (normal) hierarchy. As examples, the benchmarks are chosen as $m_{\Delta}=500$ and $1000$ GeV and the final states correspond to Eq.~\eqref{proc:sig} with $\ell=e$ (for IH) and $\ell=\mu$ (for NH), while the histograms for background correspond to Eq.~\eqref{proc:bkg12} to \eqref{proc:bkg4}. For simplicity, we show only one of the neutrino mass hierarchies in each figure of kinematic distributions and the histograms corresponding to the other not shown behave similarly to the displayed one. One can tell from Figure.~\ref{fig:dist}(a) and (b) that the number of electrons or muons for either IH or NH centers around 4, while for the SM backgrounds, the distributions reveal a much smaller electron or muon number. Since each pair of final same-sign leptons in the signal events comes from the decay of a heavy scalar, their momenta tend to be much larger than those in the background events. And as the mass of $\Delta$ increases, the peak values grow larger as well, which can be seen clearly from the distributions of leading lepton $p_{T}$ in Figure.~\ref{fig:dist}(c) and (d). Hence, one of the key features to distinguish the signal from backgrounds is multiple leptons with high $p_{T}$. Another distinctive feature of the signal also comes from $\Delta$ decay which is displayed in Figure.~\ref{fig:dist}(e) and (f), showing distributions of the invariant mass reconstructed from same-sign electrons or muons. For signal events, the histograms exhibit clear endpoints centering around the values of $m_{\Delta}$, while the background ones peak at a smaller range with either short or long tails.

\begin{table}
\centering
\begin{tabular}{|l|c|c|c|c|c|}
\hline
  & \makecell{$m_{\Delta}=500$ \\ NH} &  $\ell^{+}\ell^{-}Z/\gamma$ & $W^{+}W^{-}Z/\gamma$ & $W^{+}W^{-}jj$ & $t\bar{t}$ \\ \hline
No cuts & $1.45\times10^{-6}$ & $8.52\times10^{-6}$ & $4.68\times10^{-7}$ & $2.55\times10^{-3}$ & $7.71\times10^{-6}$ \\\hline
2 protons & $9.63\times10^{-7}$  & $9.01\times10^{-7}$ & $2.45\times10^{-7}$ & $4.88\times10^{-4}$ & $3.15\times10^{-6}$ \\ \hline
$2\mu^{+}2\mu^{-}$ & $5.86\times10^{-7}$ & $1.24\times10^{-7}$ & $1.37\times10^{-8}$ & $5.10\times10^{-9}$ & $4.04\times10^{-9}$ \\ \hline
High $p_{T}$ & $5.86\times10^{-7}$ & $1.10\times10^{-7}$ & $1.34\times10^{-8}$ & $2.55\times10^{-9}$ & $2.82\times10^{-9}$ \\ \hline
$m(\mu^{\pm}\mu^{\pm})$ & $4.69\times10^{-7}$ & $9.46\times10^{-10}$ & $1.08\times10^{-10}$ & $0$& $0$ \\
\hline
\end{tabular}
\caption{Effective cross sections of the signal process $pp\to p(\gamma\gamma\to\Delta^{++}\Delta^{--}\to\mu^{+}\mu^{+}\mu^{-}\mu^{-})p$ after each step of cutflow under NH neutrino mass spectrum with $m_{\Delta^{\pm}}=500$ GeV, and of four SM backgrounds $\ell^{+}\ell^{-}Z/\gamma$, $W^{+}W^{-}Z/\gamma$, $W^{+}W^{-}jj$ and $t\bar{t}$ events from elastic photon fusion at the 14 TeV LHC. Cross sections and masses are in units of picobarn and GeV, respectively, which are omitted in the table for simplicity.}
\label{tab:cutflow-NH4m}
\end{table}

Based on the above kinematic distributions for $4e$ events of the IH case and $4\mu$ events of the NH case, we perform event selections (Table.~\ref{tab:selection}) to achieve a better signal significance for each case. As discussed, the final two intact protons are detected with certain efficiencies (Table.~\ref{tab:protonrates}), we can regard this procedure as a step of pre-selection of the events. Then two pairs of opposite same-sign electrons ($e^{+}e^{+}e^{-}e^{-}$) and muons ($\mu^{+}\mu^{+}\mu^{-}\mu^{-}$) are required as cut-1 for IH and NH, respectively. Cut-2 is $p_{T}>50$ GeV for the leading lepton and cut-3 on the invariant mass of the pair of same-sign leptons $m({\ell^{\pm}\ell^{\pm}})$. To realize a relatively optimal selection strategy and signal significance, we apply specific selection criteria on $m({\ell^{\pm}\ell^{\pm}})$ to each benchmark point of $m_{\Delta}$. For the final states of $e^{+}e^{+}e^{-}e^{-}$ in signal region of IH, cut-3 is adopted as $m(e^{\pm}e^{\pm})\in[m_{\Delta}-25,m_{\Delta}+15]$ GeV, while for $\mu^{+}\mu^{+}\mu^{-}\mu^{-}$ final states in signal region of NH, cut-3 is $m(\mu^{\pm}\mu^{\pm})\in[m_{\Delta}-20,m_{\Delta}+20]$ GeV. As examples, Table.~\ref{tab:cutflow-NH4m} and \ref{tab:cutflow-IH4e} display cutflows of effective cross sections for the signal and background through these cuts for $m_{\Delta}=500$ and 850 GeV under NH and IH, respectively. For both cases, we can see from the tables that cut-1 on leptonic multiplicities can largely suppress the backgrounds of $W^{+}W^{-}jj$ and $t\bar{t}$ events, while cut-3 on invariant mass of same-sign lepton pair can basically filter out the other two $\ell^{+}\ell^{-}Z/\gamma$ and  $W^{+}W^{-}Z/\gamma$.

\begin{table}
\centering
\begin{tabular}{|l|c|c|c|c|c|}
\hline
  & \makecell{$m_{\Delta}=850$ \\ IH} &  $\ell^{\pm}\ell^{\pm}Z/\gamma$ & $WWZ$ & $WWjj$ & $t\bar{t}$ \\ \hline
No cuts & $1.37\times10^{-7}$ & $8.52\times10^{-6}$ & $4.68\times10^{-7}$ & $2.55\times10^{-3}$ & $7.71\times10^{-6}$ \\\hline
2 protons & $8.87\times10^{-8}$  & $9.01\times10^{-7}$ & $2.45\times10^{-7}$ & $4.88\times10^{-4}$ & $3.15\times10^{-6}$ \\ \hline
$2e^{+}2e^{-}$ & $7.47\times10^{-8}$ & $1.54\times10^{-7}$ & $1.87\times10^{-8}$ & $1.28\times10^{-7}$ & $3.04\times10^{-9}$ \\ \hline
High $p_{T}$ & $7.47\times10^{-8}$ & $1.37\times10^{-7}$ & $1.83\times10^{-8}$ & $9.95\times10^{-8}$ & $2.22\times10^{-9}$ \\ \hline
$m(e^{\pm}e^{\pm})$ & $5.54\times10^{-8}$ & $4.26\times10^{-11}$ & $1.65\times10^{-11}$ & $0$& $0$ \\
\hline
\end{tabular}
\caption{Same as Table.~\ref{tab:cutflow-NH4m} but for the signal process $pp\to p(\gamma\gamma\to\Delta^{++}\Delta^{--}\to e^{+}e^{+}e^{-}e^{-})p$ under IH neutrino mass spectrum with $m_{\Delta^{\pm}}=850$ GeV.}
\label{tab:cutflow-IH4e}
\end{table}

\begin{figure}[t]
\centering
\includegraphics[scale=0.8]{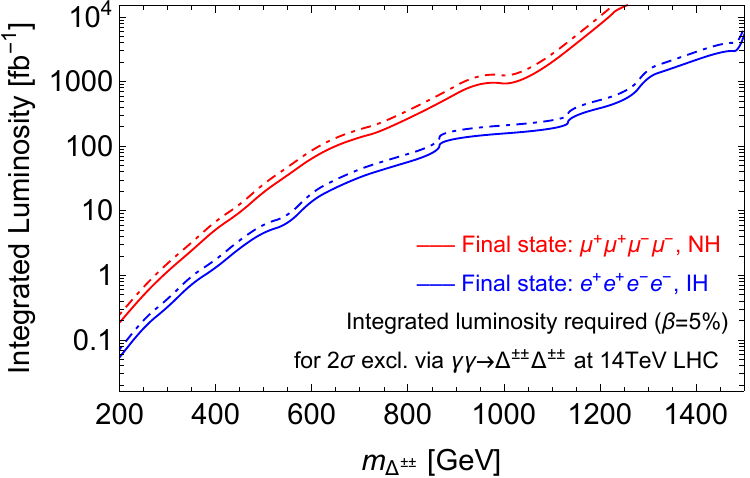}
\caption{Integrated luminosities required to reach $2\sigma$ exclusion searching for fully elastic photon fusion $pp\to p(\gamma\gamma\to\Delta^{++}\Delta^{--}\to\ell^{+}\ell^{+}\ell^{-}\ell^{-})p$ at the 14 TeV LHC. Systematic uncertainty $\beta$ is assumed as 5\%. Solid lines correspond to results using MMHT+LUXqed NLO PDF sets, while dotdashed lines correspond to results considering corrections using $\gamma$-UPC package~\cite{Shao:2022cly} as PDFs for photon fusion.}
\label{fig:Lum}
\end{figure}

\begin{figure}[t]
\centering
\begin{minipage}{0.38\linewidth}
  \centerline{\includegraphics[scale=0.575]{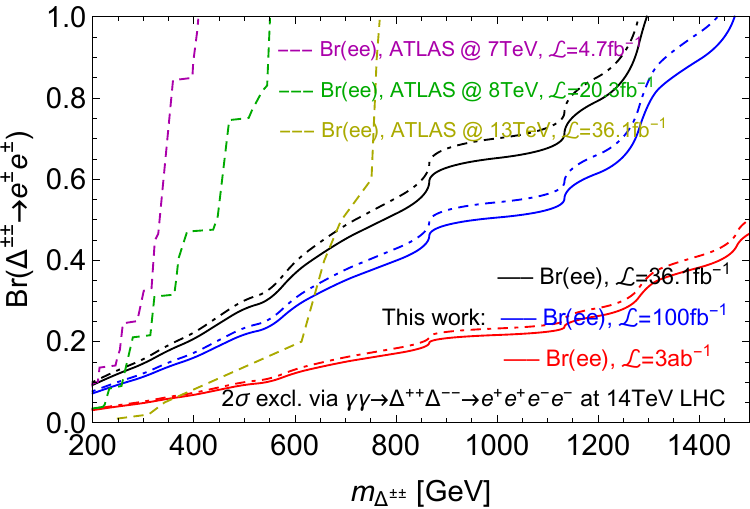}}
  \centerline{(a)}
\end{minipage}
\qquad\qquad
\begin{minipage}{0.38\linewidth}
  \centerline{\includegraphics[scale=0.575]{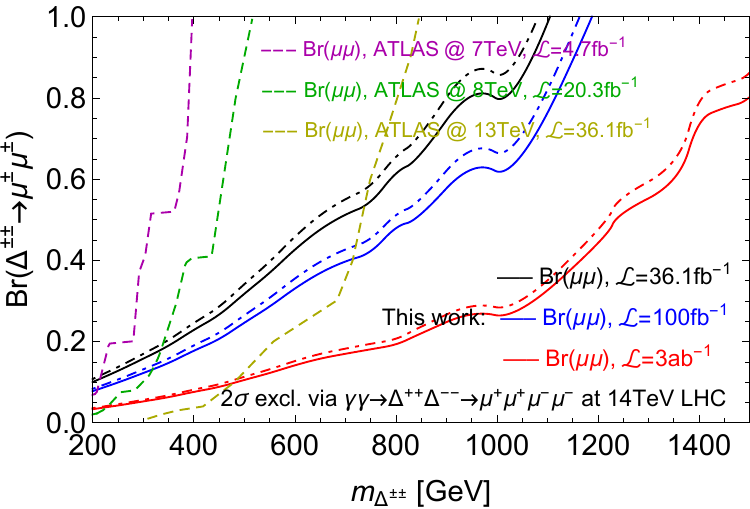}}
  \centerline{(b)}
\end{minipage}
\caption{Branching ratios that can be excluded at 95\% C.L. via searching for fully elastic photon fusion $pp\to p(\gamma\gamma\to\Delta^{++}\Delta^{--}\to\ell^{+}\ell^{+}\ell^{-}\ell^{-})p$ at the 14 TeV LHC, shown in black ($\mathcal{L}=36.1\text{fb}^{-1}$), blue ($\mathcal{L}=100\text{fb}^{-1}$) and red ($\mathcal{L}=3\text{ab}^{-1}$) solid lines. Systematic uncertainty $\beta$ is assumed as 5\%. Figure (a) and (b) are respectively for branching ratios of $\Delta^{\pm\pm}\to e^{\pm}e^{\pm}$ and $\mu^{\pm}\mu^{\pm}$. Solid lines correspond to results using MMHT+LUXqed NLO PDF sets, while dotdashed lines correspond to results considering corrections using $\gamma$-UPC package~\cite{Shao:2022cly} as PDFs for photon fusion. Dashed lines are bounds from the ATLAS experiment~\cite{ATLAS:2012hi,ATLAS:2014kca,ATLAS:2017xqs}.}
\label{fig:exclBr}
\end{figure}

Masses of the doubly charged scalar are scanned with a step of 50 GeV from 200 to 1500 GeV under both IH and NH spectra of neutrino mass. Event selections are then applied according to Table.~\ref{tab:selection} with regard to two different final states of $e^{+}e^{+}e^{-}e^{-}$ and $\mu^{+}\mu^{+}\mu^{-}\mu^{-}$, under neutrino mass spectra of IH and NH, respectively. The expected significance is finally calculated for each scanned point of $m_{\Delta}$ using the formula
\begin{align}
\alpha=S/\sqrt{B+(\beta B)^{2}}\,,
\end{align}
where $S(B)$ refers to signal (background) events after the event selections in Table.~\ref{tab:selection} and $\beta$ the systematic uncertainty. Given the parameters input for the neutrino mixing matrix in the cases of IH and NH (Eq.~\ref{pmns:IH} and \ref{pmns:NH}), we obtain the luminosities needed to arrive at $2\sigma$ exclusion limits for every scanned $m_{\Delta}$, the fitted curves of which are shown in Figure.~\ref{fig:Lum}. For luminosities of 36.1 fb$^{-1}$, 100 fb$^{-1}$ and 3 ab$^{-1}$, $2\sigma$ exclusion limits are also obtained for the branching ratios of $\Delta\to\mu\mu$ for NH and the ones of $\Delta\to ee$ for IH at each point of $m_{\Delta}$. We present in Figure.~\ref{fig:exclBr} showing contours in the parametric space of the diagonal branching ratios versus $m_{\Delta}$. As a comparison, exclusion bounds are also presented from searching for prompt same-sign lepton pairs by the ATLAS experiment~\cite{ATLAS:2012hi,ATLAS:2014kca,ATLAS:2017xqs} at 7, 8 and 13 TeV with 4.7, 20.3 and 36.1 fb$^{-1}$ of integrated luminosities, respectively.

From Figure.~\ref{fig:Lum} and \ref{fig:exclBr} one can see that sensitivities for $e^{+}e^{+}e^{-}e^{-}$ final states are better than that for $\mu^{+}\mu^{+}\mu^{-}\mu^{-}$. This performance results partly from our assumptions for NH and IH neutrino mass spectra under which the $\Delta$ decay branching ratio into electrons ($\sim47\%$) is larger than that into muons ($\sim25\%$). As $m_{\Delta}$ increases, the cross sections of elastic photon fusion production of $\Delta^{++}\Delta^{--}$ decrease rapidly (Figure.~\ref{fig:xs}), leading to lower sensitivities for branching ratios in the range of larger masses as expected. Despite this, with a collision energy of 14 TeV and integrated luminosity of 36.1 fb$^{-1}$, a $2\sigma$ exclusion limit for $m_{\Delta}$ can reach 730 GeV for Br$(\Delta\to ee)\sim47\%$ under IH spectrum and surpass the limits given by the ATLAS experiment at $m_{\Delta}\sim685$ GeV~\cite{ATLAS:2017xqs}. With higher luminosities of 100 fb$^{-1}$ and 3 ab$^{-1}$, this bound can be extended to $m_{\Delta}\sim880$ GeV and 1.5 TeV, respectively. However, considering the angular limitation of forward detectors for outgoing protons, the invariant mass of final states is bounded from above around 2.6 TeV~\cite{Duarte:2022xpm}, the best exclusion limit of 1.5 TeV can then be regarded as 1.3 TeV for the case of IH spectrum. On the other hand, with luminosities of 36.1 fb$^{-1}$ and 100 fb$^{-1}$ the results through $\mu^{+}\mu^{+}\mu^{-}\mu^{-}$ are less promising, excluding at 95\% C.L. around $m_{\Delta}\sim430$ and 520 GeV, respectively, for Br$(\Delta\to\mu\mu)\sim25\%$ under the NH spectrum. But with a higher luminosity of 3 ab$^{-1}$, the mass exclusion bound can be improved up to $m_{\Delta}\sim1$ TeV, surpassing the limit set by the ATLAS experiment at $m_{\Delta}\sim620$ GeV~\cite{ATLAS:2017xqs}.

It should be noted that development on $\gamma$-PDF and its implementation in automated event generation will improve the accuracy of phenomenological studies related to photon fusion at hadron colliders, one of which is the recently developed $\gamma$-UPC package~\cite{Shao:2022cly} which incorporates effects including hadronic survival probabilities. As a conservative estimate, if we make a correction for the photon fusion with $\gamma$-UPC by adopting 75\% of the cross sections given by MMHT+LUXqed NLO PDF sets, the exclusion mass bounds related to branching ratios are loosened but can still be comparable with current experimental limits. These corrected contours are shown in Figure.~\ref{fig:Lum} and \ref{fig:exclBr} as dotdashed lines with the same colors corresponding to each of the above luminosities.

\section{Conclusion}
We study the sensitivities for pair production of doubly charged scalars within the type-II seesaw mechanism (with a degenerate mass spectrum of the scalar components) through EPA photon fusion at the 14 TeV LHC: $pp\to p(\gamma\gamma\to\Delta^{++}\Delta^{--})p$. Two kinds of neutrino mass spectra, normal and inverted hierarchies, are considered. In the case of normal hierarchy, final states of $\mu^{+}\mu^{+}\mu^{-}\mu^{-}$ are studied as the signal process with integrated luminosities of 36.1 fb$^{-1}$, 100 fb$^{-1}$ and 3 ab$^{-1}$, the $2\sigma$ exclusion limits on $m_{\Delta}$ can reach around 430, 520 and 1000 GeV respectively, for branching ratio of $\Delta^{\pm\pm}\to\mu^{\pm}\mu^{\pm}\sim$ 25\%, which is set by adopting the best-fit values of PMNS matrix under NH spectrum. And for the case of inverted hierarchy and a largely fixed branching ratio $\Delta^{\pm\pm}\to e^{\pm}e^{\pm}\sim$ 47\%, final states of $e^{+}e^{+}e^{-}e^{-}$ are studied as signal under the same luminosities with the exclusion bounds on $m_{\Delta}$ reaching about 730, 880 and 1300 GeV, corresponding to the above luminosities. The results for exclusion bounds can also be comparable to experimental limits if we take into consideration of the correction by using $\gamma$-UPC package as PDFs for photon fusion.

\section{Acknowledgments}
We are grateful to David d'Enterria and Hua-Sheng Shao for their helpful discussions on $\gamma$-PDFs. This work is supported by the National Natural Science Foundation of China under Grant No. 12405118 and the Natural Science Foundation of Jiangsu Province under Grant No. BK20230623.


\end{document}